\documentstyle[prb,aps,epsf,twocolumn]{revtex}
\begin{document}
\draft

\title{Binding energies and electronic structures of adsorbed titanium chains on carbon nanotubes}

\author{Chih-Kai Yang,$^1$ Jijun Zhao,$^{2}$ and Jian Ping Lu$^{2}$}

\address{$^{1}$ Center for General Education, Chang Gung University, Kueishan, Taiwan 333, Republic of China \\
$^2$ Department of Physics and Astronomy, University of North Carolina at Chapel Hill, Chapel Hill, NC 27599, USA}

\maketitle

\begin{abstract}

We have studied the binding energies and electronic structures of metal (Ti, Al, Au) chains adsorbed on single-wall carbon nanotubes (SWNT) using first principles methods. Our calculations have shown that titanium is much more favored energetically over gold and aluminum to form a continuous chain on a variety of SWNTs. The interaction between titanium and carbon nanotube significantly modifies the electronic structures around Fermi energy for both zigzag and armchair tubes. The delocalized $3d$ electrons from the titanium chain generate additional states in the band gap regions of the semiconducting tubes, transforming them into metals. 

\end{abstract}

\pacs{61.48.+c, 71.15.Nc, 73.20.Hb, 71.15.Hx}

Recent experiments \cite{1,2} have offered definitive evidence that titanium atoms deposited on a single-wall carbon nanotube are capable of forming continuous wires. Compared with other metals such as gold, palladium, iron, aluminum, and lead, Ti atoms have stronger interactions with carbon nanotube, higher nucleation, and a lower diffusion rate. Nanotubes coated with a Ti buffer layer also have a better chance of forming continuous wires made of other metals. Thus, they function as templates for growing metal nanowires, which may be suitable for various applications in nanoscale materials and devices. A nanotube with adsorbed materials may also significantly changes its physical properties, providing useful means for manipulating electronic transport for nanoelectronic devices. Moreover, understanding the interaction between titanium and nanotube will be helpful in reducing the contact resistance between nanotubes and metal leads, which is a critical problem in the nanoelectronic applications \cite{3,4,5}.

In this article we present {\em ab initio} results on several SWNTs with adsorbed metal chains. We demonstrate theoretically that Ti is favored energetically over Au and Al, two popular materials for nanowires, for the formation of a continuous chain on the carbon nanotubes. Detailed analysis shows significant modification of electronic states of the carbon nanotube after incorporating a Ti chain. In particular, semiconducting zigzag SWNTs transform into metals upon adsorption of Ti chain.  

Electronic structure calculations and geometry optimization were performed using the Vienna {\em ab initio} simulation package VASP, \cite{6,7,8} a plane-wave  pseudopotential program. Ultrasoft pseudopotentials were used with a uniform energy cutoff of 210 eV. There were 31 uniformly distributed {\bf k} points along the nanotube symmetry axis ${\Gamma} X$ for both the pure SWNT and the SWNT deposited with a metal chain. Ion positions were relaxed into their instantaneous ground states using a conjugate-gradient algorithm. During each ionic step, a maximum of 120 self-consistent cycles were executed and the Hellmann-Feynman forces were calculated. We included two zigzag SWNTs, (10,0) and (14,0), and two armchair ones, (6,6) and (8,8) in our calculations.  Different initial configurations of the adsorbed metal chains have been used in our calculation to obtain the lowest-energy structures.

We have optimized the equilibrium structures and calculated total energies for each pure SWNT, free-standing metal chain, and metal-deposited SWNT. The binding energy of the metal chain on SWNT is defined as the difference between the total energy of the metal-adsorbed SWNT and the sum of total energies of the individual SWNT and free-standing metal chain. For both Al and Au, we find interaction between metal atoms and the carbon nanotube is weak. The binding energy obtained for aluminum chain on armchair (6,6) SWNT is 0.52 eV per metal atom and that for gold on (6,6) SWNT is only 0.25 eV/atom. Similar results are found in the case of zigzag (10,0) SWNT for Al and Au. The weak interactions between Al and carbon nanotube are consistent with a recent {\em ab initio} calculation of the nanotube on aluminum surface \cite{5}. The low binding energies ($\leq 0.5$ eV) suggest that the interaction between Al, Au metal and carbon nanotube is most likely physisorption. Thus, a low barrier against the diffusion of Al, Au adatoms on SWNTs is expected. Between these two metals, Al has a higher binding energy than Au, confirming part of the suggested order of binding energy in Ref. 2. With only one valence electron in the $3p$ shell Al is more likely to have stronger interaction with the nanotube than Au, which has nearly filled outer $sd$ shells. 

{\begin{table} {Table I. Binding energy per metal atom $E_b$ and equilibrium metal-nanotube distance $d$ for adsorbed Ti chains on SWNTs. The tube radiuses are also presented.}
\begin{center}
\begin{tabular} {ccccc} 
  Tube             &  (10,0)  &  (6,6)  &  (8,8)  &  (14,0)  \\ \hline
  Radius ({\AA})   &  3.91    &   4.07  &  5.42   &  5.48    \\
   $d$ ({\AA})     & 2.23     &  2.09   &  2.11   &  2.24    \\
   $E_b$ (eV)      &  1.94    &   2.04  &  1.62   &  1.97    \\ 
\end{tabular}
\end{center}
\end{table}}

By contrast, the binding energy between Ti and the nanotube is much greater than the cases of Al and Au, as shown in Table I. It is known that $3d$ orbitals in titanium are rather delocalized and function as valence orbitals \cite{9}. Thus, the unfilled $3d$ shell of Ti may lead to certain hybridization between $3d$ orbitals of Ti and $2p$ orbitals of C, resulting in a stronger interaction between them. Indeed, the calculated binding energies of the adsorbed Ti chains range from 1.62 to 2.04 eV per titanium atom for the SWNTs studied. As shown in Table I, the binding energy is not sensitive to either tube size or chirality. Such higher adsorption energy favors nucleation of Ti atoms and is capable of forming continuous wires, as shown by the recent experiments. \cite{1,2} According to an empirical estimate \cite{10} the diffusion activation energy $E_{diff}$ is about a quarter of the binding energy and the diffusion rate is proportional to $exp(-{E_{diff}}/{k_B}T)$, where $k_B$ is the Boltzmann constant and $T$ is the absolute temperature. If we follow this scheme and calculate diffusion rates for Ti, Al, and Au, we will see that Ti has a much smaller probability than Al and Au to diffuse. A 1.5 eV difference in binding energy per atom, for instance, makes Ti harder to diffuse by an order of $10^{-7}$ at room temperature. Thus, Ti atoms are much more favored for forming the stable structures coated on the surface of carbon nanotubes than Al or Au atoms \cite{1,2}. 

\begin{figure}
\vspace{0.1in}
\centerline{
\epsfxsize=3.0in \epsfbox{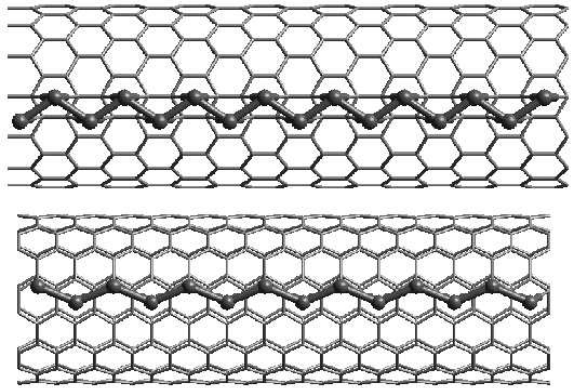}
}
\caption{Geometric structures of adsorbed Ti chains on (14,0) (top) and (8,8) (bottom) tubes. Along the tube axis, each unit cell contains two Ti atoms for  (14,0) tube and one Ti atom for (8,8) tube. Several unit cells are shown for purpose of visualization. Distances between the two neighboring Ti atoms in the Ti chain are 2.6 {\AA} for (14,0) and 2.7 {\AA} for (8,8) tubes.}
\end{figure}

The fully relaxed atomic structures of Ti chains adsorbed on carbon nanotubes are illustrated in Figure 1 for (8,8) and (14,0) SWNT respectively. Along the tube axis, per unit cell there are two Ti atoms for the zigzag tube and one for the armchair tube. The deformation of carbon nanotube due to titanium adsorption is relatively small. The deviation from circular tube as measured by the aspect ratio ($r_{max}/r_{min}$) \cite{11} is less than 2$\%$ for all SWNTs studied (see also Figure 2(c) and Figure 3(c)). Distances between the two neighboring Ti atoms in the adsorbed metal chain are about 2.6 {\AA} for zigzag tubes and 2.7 {\AA} for armchair tubes, which are smaller than the bulk interatomic distance 2.89 {\AA} and comparable to the bond length in titanium clusters \cite{9}.

\begin{figure}
\vspace{-0.6in}
\centerline{
\epsfxsize=3.0in \epsfbox{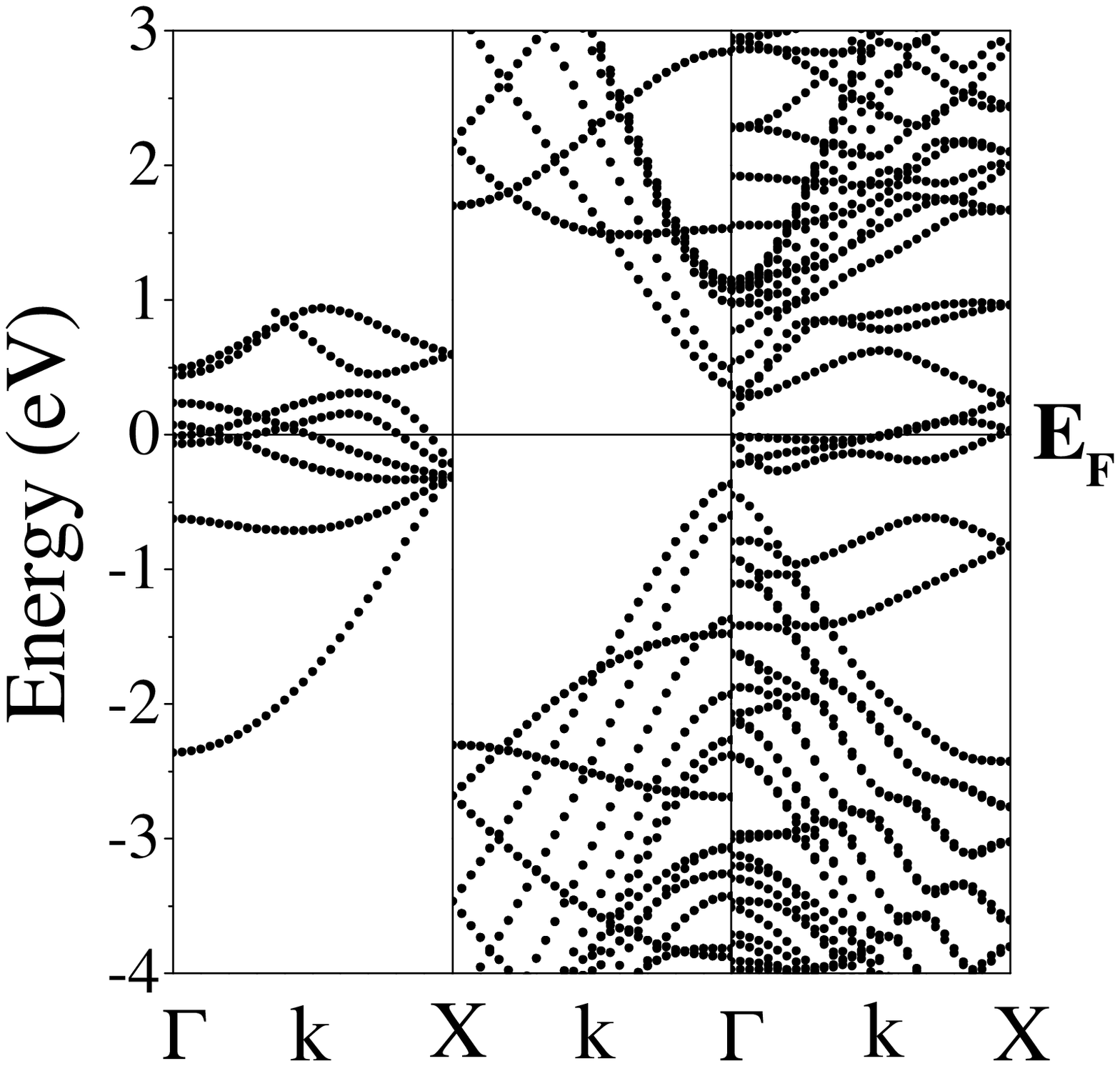}
}
\vspace{-1.65in}
\centerline{
\epsfxsize=3.0in \epsfbox{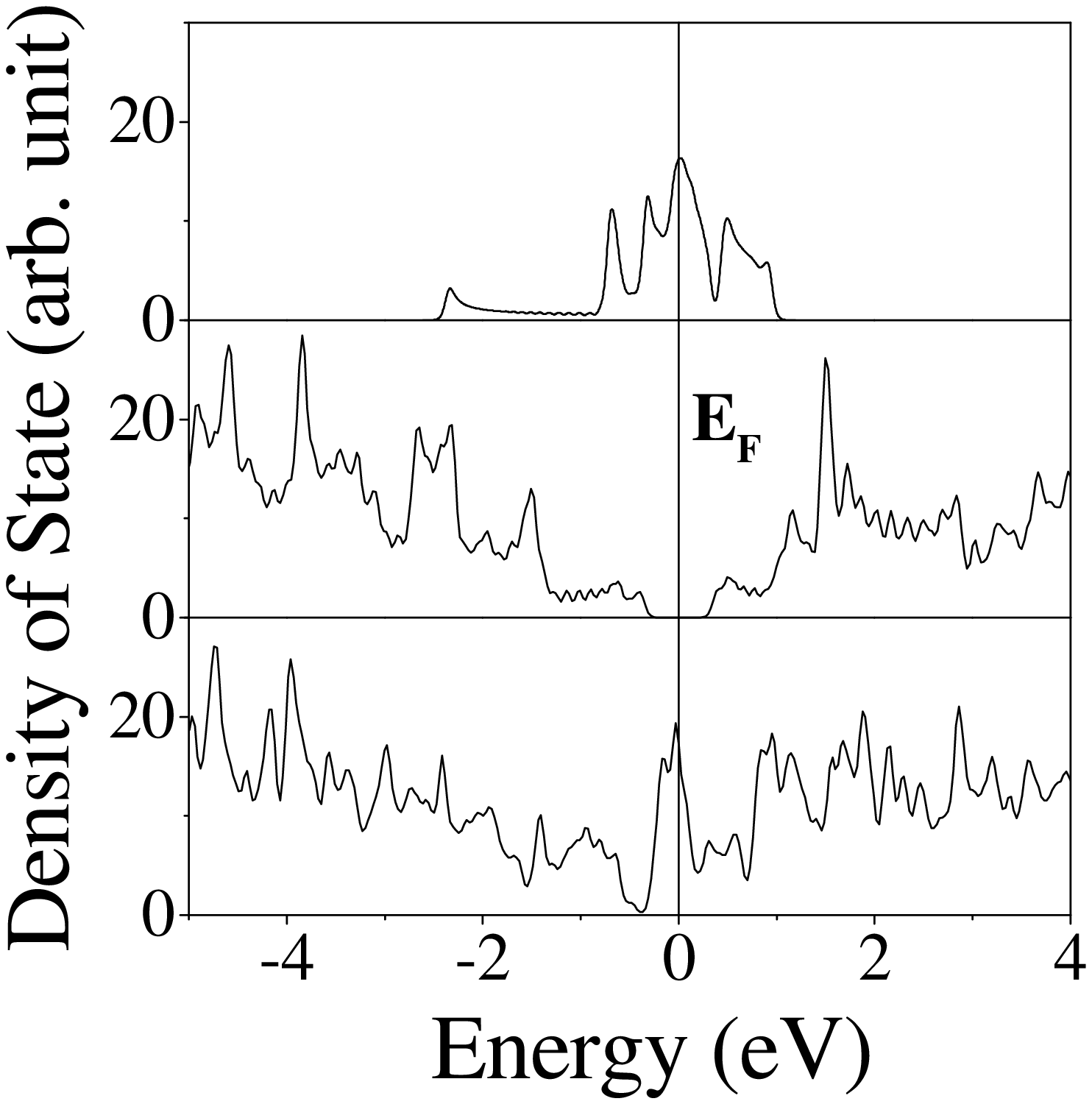}
}
\vspace{-1.1in}
\centerline{
\epsfxsize=2.0in \epsfbox{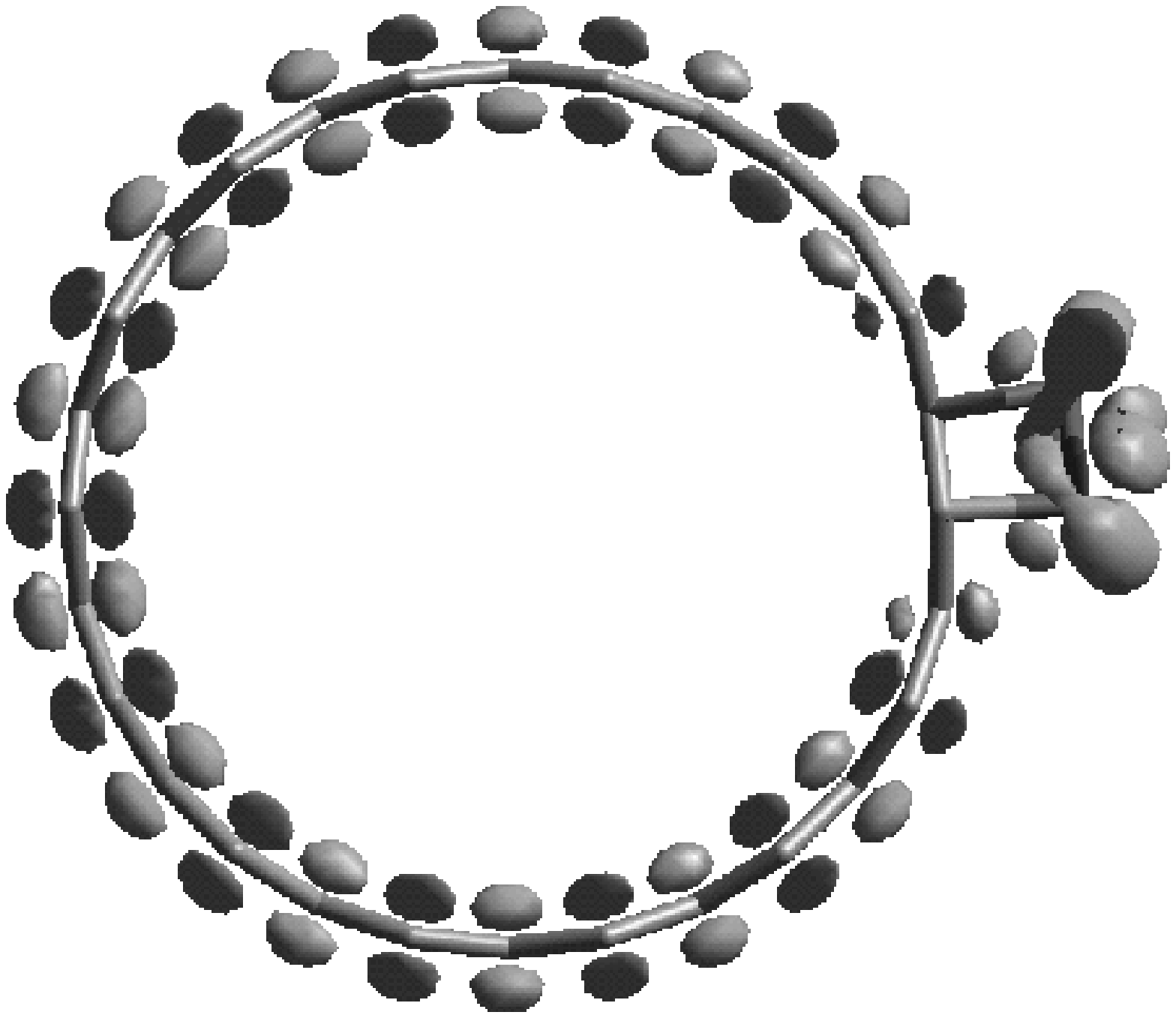}
}
\caption{(a) Band structure for a free standing Ti chain (left), a pure (14,0) nanotube (middle), and (14,0) SWNT with an adsorbed Ti chain (right); (b) Corresponding Density of states from top to bottom. A Fermi broadening of 0.03 eV is used. (c) Isosurface for the electron density of the four closest bands around the Fermi level for the Ti-adsorbed (14,0) tube. It clearly shows that the conduction electrons near the Fermi level are delocalized and distributed over both Ti chain and carbon nanotube. }
\end{figure}

The electronic band structure for Ti-adsorbed (14,0) nanotube is presented in the right part of Figure 2(a). For comparison the middle part of the diagram shows the band structure for the pure (14,0) SWNT, and the left part shows the band structure of the free-standing Ti chain. The pure (14,0) SWNT is a semiconductor with small gap. The adsorbed Ti chain, however, creates several new states in the band gap region, effectively transforms the semiconducting SWNT into a metal. This metallic behavior is also seen in the bottom of Figure 2(b), which displays the density of states (DOS) for the Ti-adsorbed SWNT, as opposed to the middle of the figure showing a band gap for the pure (14,0) and the upper one showing the DOS for the titanium chain. These newly generated bands in the gap region are apparently related to the delocalized states from the adsorbed Ti chain. Moreover, as shown in Figure 2(a), the band structure of Ti-adsorbed (14,0) nanotube is not a simple superposition of those of individual nanotube and titanium chain. The significant modification on the electronic states pertaining to titanium and carbon nanotube implies certain charge transfer between them.

To further explore the interaction between titanium and carbon nanotube and understand the delocalized electronic states around the gap region, Figure 2(c) shows the isosurface for the electron density of four bands that closest to the Fermi level. We find that the electrons from these bands are delocalized and distributed on both titanium and carbon atoms. From the band structures shown in Figure 2(a), we know that the bands around Fermi level are originally from titanium chain. Therefore, it is not surprising to find certain portion of electron density at the titanium sites. Our detailed analysis on the partial electron local density of states find that the DOS at Fermi level are composed of $3d$ electrons from titanium and $2p$ electrons from carbon. The metallization of semiconducting nanotubes caused by charge transfer from metal atoms to SWNT has also been found in the alkali-metal intercalated SWNTs \cite{12}. The delocalized nature of the conduction electrons and the charge transfer imply a possible good contact between titanium and carbon nanotube. This finding might be helpful in reducing the contact resistance between nanotube and metal leads \cite{3,4}. We have also carried out detailed analysis on (10,0) nanotube with adsorbed titanium chain. Similar results are obtained. Thus, we believe these behaviors are general for all semiconducting zigzag nanotubes.

Shown in the middle part of Figure 3(a) is the band structure of the pure (8,8) nanotube. At the Fermi level, there is a well-known band crossing at about two thirds of the distance between ${\Gamma}$ and $X$, making it a metallic nanotube \cite{12}. Again, the electronic structures of carbon nanotube have been significantly modified by the titanium chain. The band structure of the Ti-adsorbed (8,8) nanotube is not simply the superimposition of those of individual nanotube and titanium chain (see Figure 3(a)). It is interesting to note that the adsorbed Ti chain induces a small gap of about 0.14 eV. As shown in the right part of Figure 3(a), an almost dispersionless band lies just above the Fermi level, while another equally flat band lies lower in energy, both implying large effective mass for the electron. Such opening of a small gap has been reported for the Cu-adsorbed SWNT \cite{13}. Due to the presence of the adsorbed chain, the mirror symmetry of the armchair nanotube is broken and degeneracy of the ${\pi}$ and ${\pi}^{*}$ bands is removed \cite{14}. After a 0.03 eV Fermi broadening, a low but finite value of DOS at the Fermi level is found in the bottom of Figure 3(b) for the Ti-adsorbed nanotube with a small pseudogap. Similar effect has been found in Ti-adsorbed (6,6) tube, where the titanium-induced pseudogap is smaller ($\sim$ 0.1 eV). 

\begin{figure}
\vspace{-0.6in}
\centerline{
\epsfxsize=3.0in \epsfbox{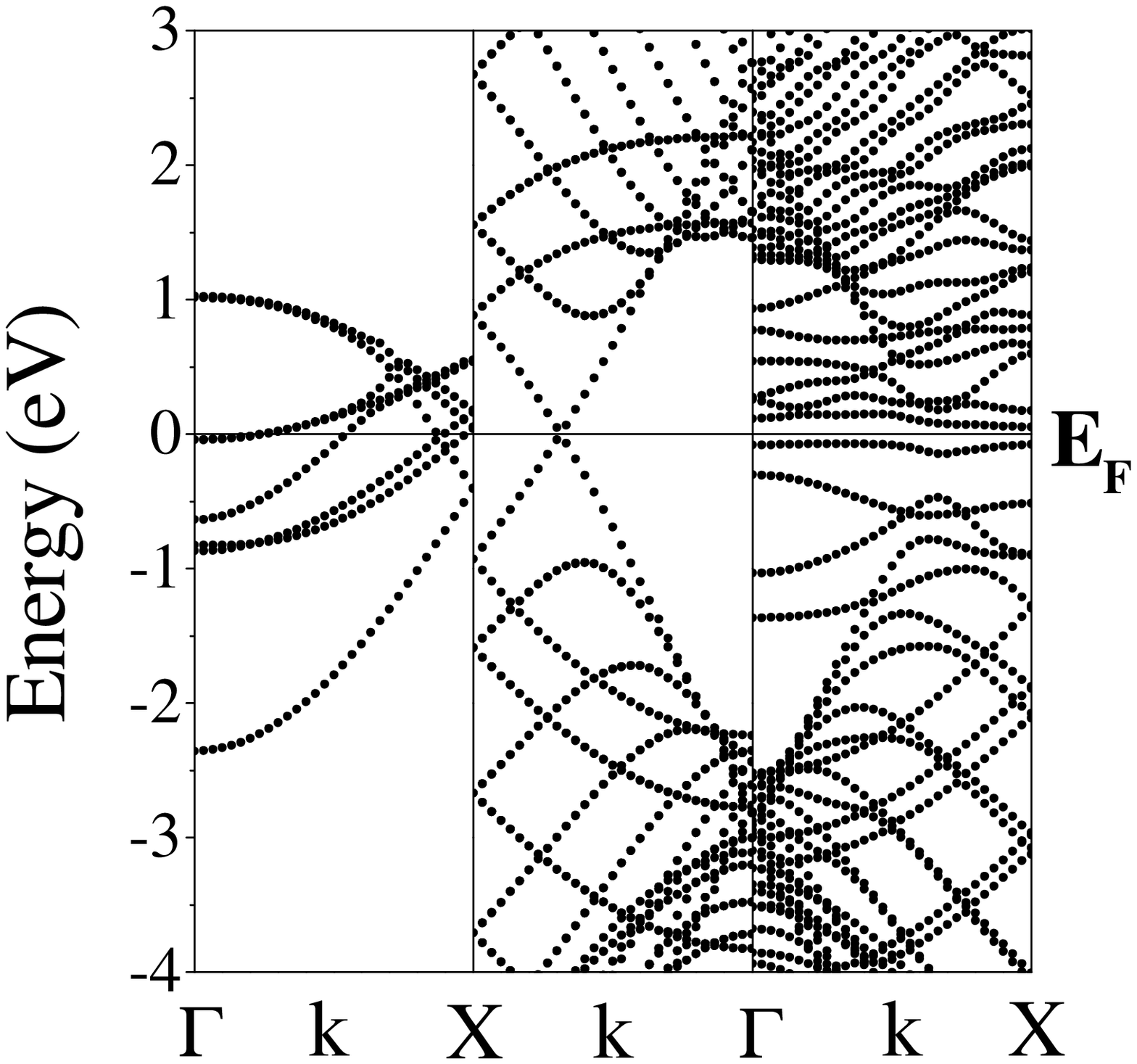}
}
\vspace{-1.65in}
\centerline{
\epsfxsize=3.0in \epsfbox{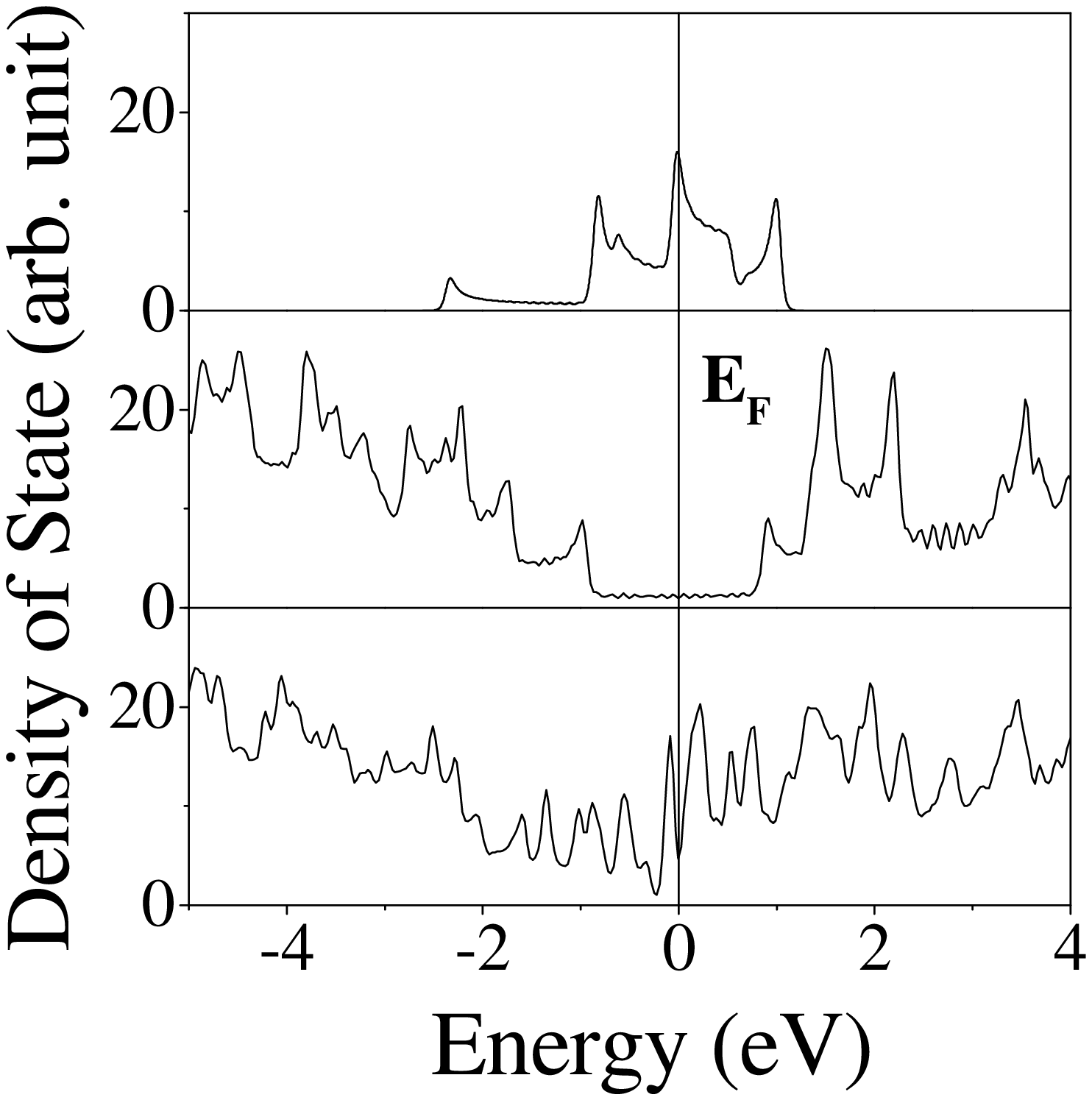}
}
\vspace{-1.1in}
\centerline{
\epsfxsize=2.0in \epsfbox{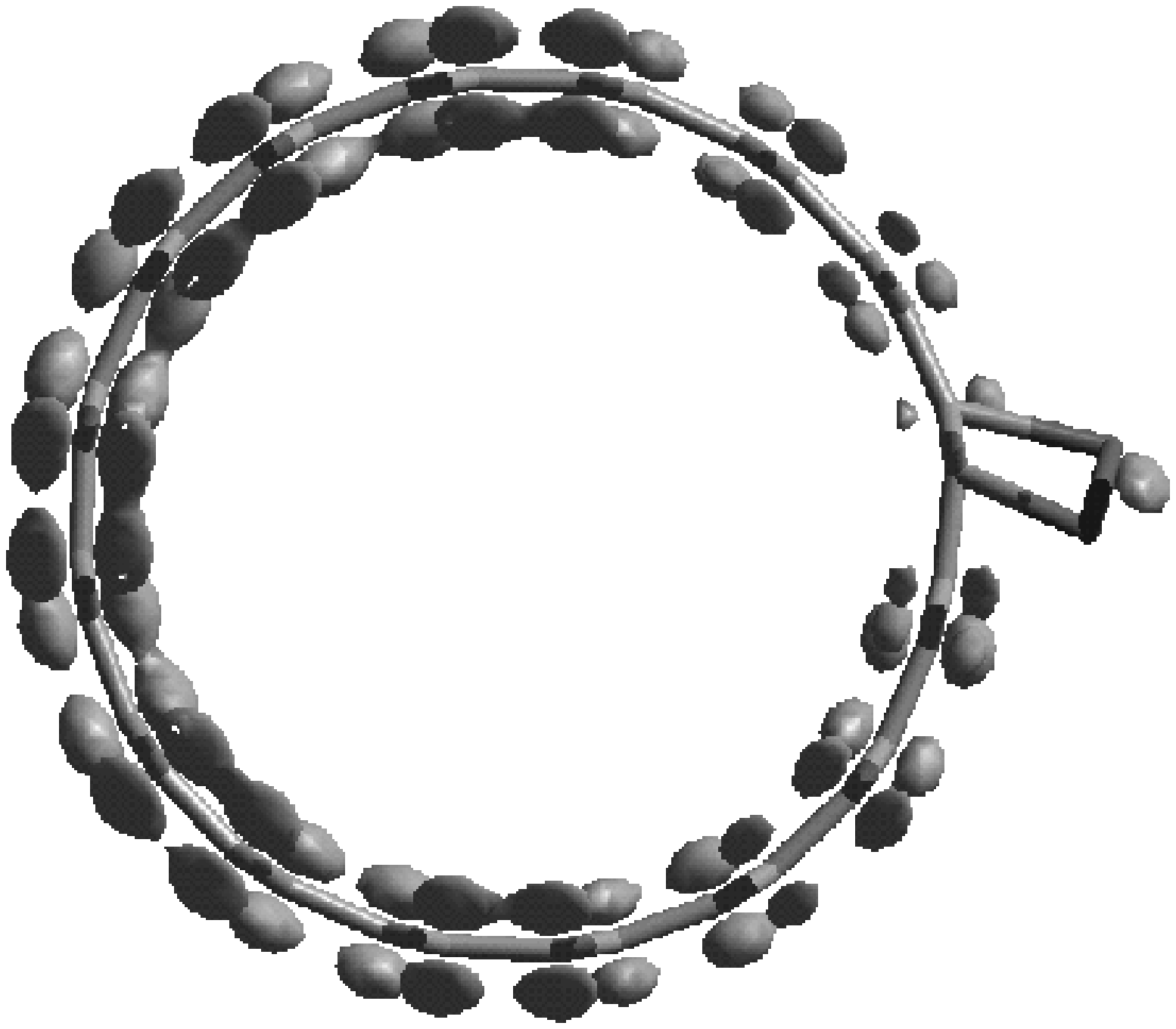}
}

\caption{The same plot of those of Figure 2 but for (8,8) tube.}
\end{figure}

Analysis of charge density for the bands around the Fermi level shows that they are mostly located on the carbon nanotube (see Figure 3(c)). Moreover, the circular symmetry of electron density distribution is disturbed by Ti chain, which leads to reduced electron density on the carbon sites near the Ti atoms. It is worthy to note that there is much less electron density on titanium in the case of armchair (8,8) tube than that in the zigzag (14,0) tube. Such difference might be understood by the existence of conduction bands at the Fermi level in the pure armchair nanotube, which leads to more electron density distribution on carbon atoms. Nevertheless, in both cases there are significant charge transfers from the Ti delocalized $3d$ electrons to carbon nanotube.

In summary, we have performed first-principles calculations on metal-adsorbed single-wall carbon nanotubes. We conclude that Ti is much more likely than Al and Au to form continuous chain on the surface of SWNTs due to much higher binding energies. Al-adsorbed SWNTs in turn have higher binding energies than the Au-adsorbed ones. The delocalized $3d$ electrons from the titanium chain partially transfer to carbon nanotube and generate additional states around the Fermi level. The band structures analysis shows that pure zigzag nanotubes transform from semiconductors into metals with the adsorption of Ti atoms, while armchair nanotubes may open up a pseudogap due to the symmetry breaking. A novel type of functional nanostructures can be anticipated with titanium adsorbed carbon nanotubes. The modification of electronic properties by the adsorbed Ti atoms also provides a useful means for a new generation of nanoelectronic devices.

This work was supported by the National Science Council of Taiwan, the Republic of China, under contract number NSC 90-2112-M-182-005 and NASA Ames Research Center. The authors gratefully acknowledge computational support from the North Carolina Supercomputer Center and the National Center for High-Performance Computing of ROC.

\end{document}